\input amstex.tex
\documentstyle{amsppt}

\magnification\magstep1

\topmatter
\title A gerbe obstruction to quantization of fermions on odd dimensional
manifolds with boundary \endtitle
\author Alan Carey and Jouko Mickelsson \endauthor
\date December 1, 1999; revised February 23, 2000 \enddate
\affil Department of Mathematics, University of Adelaide, Adelaide SA
5005, Australia, e-mail acarey\@ maths.adelaide.edu.au.  Theoretical
Physics, Royal Institute of Technology, Stockholm SE-10044, Sweden,
e-mail jouko\@ theophys.kth.se \endaffil

\endtopmatter
\NoRunningHeads
\document
\baselineskip=18pt
\NoBlackBoxes

\define\Ut{\hat U_{res}}

\redefine\a{\alpha}

\redefine\b{\beta}

\redefine\g{\gamma}

\define\tr{\text{tr}}

ABSTRACT We consider the canonical quantization of fermions on an odd
dimensional manifold with boundary, with respect to a family of elliptic
hermitean boundary conditions for the Dirac hamiltonian. We show that there
is a topological obstruction to a smooth quantization as a function of the
boundary conditions. The obstruction is given in terms of a gerbe and its
Dixmier-Douady class is evaluated.

\bf Mathematics Subject Classification (2000): \rm 81T50, 58B25, 19K56

\bf Key words: \rm Gerbe, Hamiltonian quantization, Dirac boundary
value problems 

\vskip 0.5in

0. INTRODUCTION

\vskip 0.3in

In this paper we study the Hamiltonian quantization of massless fermions on a
compact odd-dimensional manifold $X$ with boundary $Y=\partial X.$ 
Field theories on manifolds with boundary
arise in several situations including
gravitation on odd dimensional
anti-de Sitter spacetimes \cite{W}. Our aim is to
investigate topological obstructions (or anomalies)
arising from non-trivial topology in the boundary conditions. We assume
that a
Riemannian metric and spin structure is given on $X.$ The Dirac
field might also be coupled to a Yang-Mills potential but
in this
paper we shall concentrate only on the problem of the dependence of the
canonical quantization on the boundary conditions.

The obstruction to canonical quantization arises in the following
way.  We have a family of hermitean Dirac hamiltonians in a 1-particle space $H.$
Let $G$ be the parameter space for the family. In quantization the
1-particle space $H$ is replaced by a Fock space $\Cal F_g$ (with
$g\in G$) and the 1-particle hamiltonian $D_g$ by a second quantized
hamiltonian $\hat D_g.$  If the hamiltonians do not have any zero
modes this does not cause any problems. The Fock space associated to
the parameter $g$ is defined as the representation space for the
algebra of canonical anticommutation relations (CAR) with a vacuum
vector ('Dirac sea') corresponding to the polarization of the
1-particle space $H=H_+(g)\oplus H_-(g)$ to positive and negative
eigenmodes of the hamiltonian $D_g.$ 

In case of zero modes the above construction is not continuous as a
function of the parameter $g$ since $H_{\pm}(g)$ is not a continuous
function of $g.$  This is a familiar problem in the hamiltonian
quantization of chiral fermions and its resolution is related to the 
appearance of Schwinger terms in the gauge current algebra which form 
an algebraic obstruction to gauge invariant quantization of chiral
fermions. Topologically the obstruction can be understood in terms of
the structure of the space $\Cal A/\Cal G,$ the space of vector
potentials modulo gauge transformations.   

In the case at hand the space $\Cal A/\Cal G$ is replaced by the
parameter space $G$ for boundary conditions. The spectrum of the
hamiltonian $D_g$ depends on the boundary conditions and in the case
of massless fermions zero modes will occur.  

It turns out that one can avoid a detailed analysis of the zero mode
set of the family of hamiltonians (which in general is a complicated
task). Instead, the problem of constructing representations of the CAR 
algebra can be rewritten as the problem of prolonging a principal
bundle $\Cal P$ over $G,$ with structure group $U_{res},$ to a principal
bundle with the structure group $\hat U_{res}.$ Here $U_{res}$ is the
group of implementable Bogoliubov automorphisms of the CAR algebra and 
$\hat U_{res}$ is its central extension. The bundle of Fock spaces can
be viewed as an associated bundle to the prolongation, through a
representation of $\hat U_{res}$ in a Fock space.     
 
The topological obstruction to the bundle prolongation is a
three-cohomology class on the base $G,$ the Dixmier-Douady class.  In
the case of chiral fermions this has been discussed before in [CMM1]. 
In the present paper we shall compute the DD class on the parameter
space $G$ and we show that it indeed gives a nonzero obstruction to a
continuous second quantization of fermions. 

It turns out that the parameter space $G$ can be naturally identified
as the group of unitary operators $g$ in a fixed Hilbert space such
that $g-1$ is a trace class operator. The cohomology of this group is 
known. It has one generator in each odd dimension and no generators in 
even dimensions. An explicit representation for the odd generators is
given by the Wess-Zumino-Witten type differential forms $\text{tr}\,
(dg g^{-1})^{2k+1}$ on the group manifold. We are interested only on
the form with $k=1$ and we shall prove that this (when properly
normalized) is the Dixmier-Douady class for the bundle prolongation. 
  
The plan of the paper is as follows. In section 1 we shall describe
the parametrization of self-adjoint elliptic boundary conditions for
Dirac operators in odd dimensions and recall some facts about
canonical quantization. In section 2 we shall go through some
technicalities  related to local triviality of the relevant bundles
and explain in more detail the relation between canonical quantization and
the principal bundle $\Cal P.$  In section 3 we first explain general 
aspects of gerbes and their DD class and we give two different 
constructions for a universal $U_{res}$ bundle (bundle with
contractible total space) over the parameter space $G.$ These will be
needed in section 4, where we finally compute the DD class for the
bundle prolongation and the associated gerbe. 

\vskip 0.3in
1. BOUNDARY CONDITIONS AND FOCK SPACES

\vskip 0.3in

We consider Dirac operators on an odd dimensional compact manifold $X$ with
boundary $Y.$ As an additional technical assumption we require that the metric
becomes a product metric near the boundary and that the normal
component of the gauge potential $A$ and the normal derivatives of $A$
vanish at the boundary. Then near
the boundary the Dirac equation can be written as
$$ic_t \partial_t \psi= -h_Y \psi,$$
where $h_Y$ is the Dirac operator on the boundary, $t$ is a local
coordinate in the normal direction
and $c_t$ denotes Clifford multiplication by the
unit normal in the $t$ direction. The operator $h_Y$ anticommutes
with Clifford multiplication by $c_t$.

In order that the Dirac operator in the bulk $X$ becomes an elliptic
Fredholm operator we use Atiyah-Patodi-Singer type boundary
conditions. These elliptic boundary conditions are labelled by
projection operators $P$ on the boundary Hilbert space $L^2(Y, S)$
(where $S$ denotes the combined spin and gauge vector bundle) such that the
difference $P- P_+$ is a trace-class operator. Here $P_+$ is the
projection  on to the
spectral subspace on which $h_Y$ is positive.
 The boundary condition is then written as
$$P\psi|_Y =0.$$
The precise estimate on $P-P_+$ is actually not very critical in the
following. Often one requires that $P-P_+$ is a smoothing
operator. This has advantages when studying the analytical properties
of Dirac determinants. However, for analyzing topological properties
of the family of Dirac operators parametrized by such boundary
conditions it turns out to be more convenient to work in the
trace-class setting.

In order that $D_P$ is hermitean we must have
$$ P c_t = c_t(1-P).$$
We shall use the parametrization of the boundary conditions studied in
[S]:   For each boundary projection $P$ above  we 
think of $P(L^2(Y,S))$ as the graph of an unitary operator 
$$T: S^+ \to S^-,$$
where $S^{\pm}$ are the eigenspaces of the chirality operator $c_t$ on
the boundary. If $T_0$ is the unitary operator corresponding to the
projection $P_+$ then $K=T T_0^{-1}$ should differ from the identity
operator in $L^2(Y, S^-)$ by a trace-class operator.

Denote by $G$ the group of all unitaries $g$ in $L^2(Y,S^-)$ such that
$g-1$ is a trace-class operator. Thus $G$ is the parameter space of
the elliptic hermitean boundary conditions for the family of Dirac
operators $D_P$ and henceforth we write $D_g$, $g\in G$
to denote elements in  this family.

For each $g\in G$ we would like to produce a fermionic Fock space
$\Cal F_g$  carrying an irreducible quasi-free representation of the
canonical anticommutation relations (CAR) and a compatible action of the
second quantized Dirac operator $\hat D_g.$ Here quasi-free means that
there is a vacuum vector $|0> \in \Cal F_g$ such that
$$a(v)|0>=0= a^*(u) |0>$$
for all $v\in H_+$ and $u\in H_-$ where $H_+\oplus H_-$ is the
polarization of the 1-particle Hilbert space $H=L^2(X, S)$ to positive
and negative frequencies with respect to the Dirac operator $D_g.$ The
CAR algebra is generated by the relations
$$a^*(u) a(u') + a(u') a^*(u) = <u',u>$$
for $u,u'\in H$ with all other anticommutators zero and the Hilbert
space inner product $<\cdot,\cdot>$ is antilinear in the first argument.

Note that in general one cannot expect that there would be a
continuous choice $|0> =|0>_g$ of vacuum vectors parametrized by the
boundary conditions $g\in G;$ the vacuum line bundle (in case it is
well-defined) could be nontrivial. In fact, we shall show that there
is a more serious problem:  There is a topological obstruction to the
above quantization of fermions parametrized smoothly by elements of
$G.$

The obstruction appears as follows. In order to define the quasi-free
representation of the CAR algebra we need a polarization $H=H_+(g)\oplus
H_-(g)$ of the 1-particle space $H.$ This polarization should be a
continuous function of the boundary condition $g\in G.$ Furthermore,
each  Fock space $\Cal F_g$ defined by the polarization should contain
the vacuum vector for the Dirac operator $D_g=ic_t \partial_t
+h_{Y,g},$
 that is, the vacuum defined
by the splitting to positive and negative
spectral subspaces of $D_g$. This requirement
is equivalent to the condition that the plane $H_+(g)$ lies in the
infinite-dimensional Grassmannian $Gr_g$ consisting of all closed
subspaces $W\subset H$ such that the difference $P_g -Q_g$ is Hilbert-Schmidt;
here $P_g$ is the orthogonal projection onto $H_+(g)$ and $Q_g$ is the
orthogonal projection onto the positive
spectral subspace of
$D_g$.

We have now a bundle of infinite-dimensional Grassmannians
$\{Gr_g\vert\ g\in G\}$ (modelled
by the ideal of Hilbert-Schmidt operators) over the base $G$.
We still need
to show that this bundle is defined in terms of local trivializations
and smooth transition functions. Once this is done we will
show that the  potential
obstruction to quantization is the topological nontriviality of this
bundle.

\proclaim{Proposition 1}
The  bundle $Gr$ is smoothly locally trivial.
\endproclaim
The proof is given in appendix 1.

\vskip 0.3in
2. THE OBSTRUCTION TO CANONICAL QUANTIZATION

\vskip 0.3in
A  closer examination reveals that the discussion in Section 1
is not
quite accurate in the sense that often one would be satisfied 
with a 
determination of the CAR algebra representation without an explicit choice
of the vacuum vector.

To explain this let us
denote by $U_{res}$ the group of unitaries $T$ in a complex polarized
Hilbert space $H=H_+\oplus H_-$ such that
the commutator
 $[P_+, T]$ is Hilbert-Schmidt; here $P_+$ is the orthogonal
projection onto $H_+.$
This group acts naturally and transitively on the Grassmannian $Gr(H_+)$
consisting of
closed subspaces $W\subset H$ such that $P_W -P_+$ is Hilbert-Schmidt,
where $P_W$ is the orthogonal projection onto $W.$

Over $Gr(H_+)$ there
is a canonical complex line bundle $DET.$ When the
 projection from $W$
to $H_+$ has Fredholm index equal to zero, the fiber at $W$ is the set of
equivalence classes $[q,\lambda],$ where $q:H_+ \to W$ is an isomorphism
such that $P_+ q -id_{H_+}$ is trace-class and $\lambda\in \Bbb C.$  The
equivalence is defined by $(q,\lambda) \sim (qt^{-1}, \lambda \text{det}
\, t),$ where $t:H_+ \to H_+$ is an isomorphism with $t-1$ trace-class;
for details, see [PrSe].

A central extension $\hat U_{res}$ of $U_{res}$ acts in the total space
of $DET,$  [PrSe]. This extension acts unitarily in the fermionic Fock space
corresponding to the given polarization $H=H_+\oplus H_-.$

If a representation of the CAR is given with respect
to a polarization $H=H_+\oplus H_-$ then the set of (normalized) Fock
vacua is a $\hat U_{res}$ orbit through some fixed vacuum $|H_+>$ defined
by the polarization.  The orbit of $H_+$ under the $U_{res}$ action is the
infinite-dimensional Grassmannian $Gr(H_+)$ and the $\hat U_{res}$ orbit
in the Fock space is the set of vectors of unit length in the canonical
determinant bundle $DET$ over $Gr(H_+).$

In the case of a family of Grassmannians,
 the construction  of the family
of CAR representions can now be formulated as the problem of prolonging the
Grassmannian bundle to a bundle with fiber equivalent to the determinant
bundle $DET.$ There is an illuminating alternative formulation of the
prolongation  problem which we shall now describe.

The Grassmannian $Gr(H_+)$ is a homogeneous space $U_{res}/(U(H_+)\times
U(H_-)).$ The topology of the block diagonal unitary group $U(H_+)\times
U(H_-)$ is trivial by Kuiper's theorem. Thus $U_{res}$ contracts to
$Gr(H_+).$ It follows that the prolongation of a $U_{res}$ bundle over
some base manifold to a $\hat U_{res}$ bundle is equivalent to the problem
of prolonging the associated Grassmann bundle to a bundle with model fiber
$DET.$ The relevance of the $\hat U_{res}$ bundle in Fock space quantization
is immediate: selecting a model Fock space with a $\hat U_{res}$ action,
one can construct a bundle of Fock spaces as an associated vector bundle to
a given $\hat U_{res}$ bundle.

To close the circle, we note that starting from the given
Grassmanian bundle over the parameter
space $G$ one constructs a natural $U_{res}$ bundle $\Cal P$ such that the
Grassmannian
bundle is recovered as an associated bundle. If $H=H_+\oplus H_-$ is any
fixed polarization then the fiber of $\Cal P$ at $g\in G$ consists of all unitaries
$h$ in $H$ such that $h\cdot H_+ \in Gr_g.$

Instead of looking at the specific construction of the $U_{res}$
bundle over the family of boundary conditions $G$ we can extend this
to another universal construction over the space of all (bounded)
self-adjoint Fredholm operators $\Cal F_*$ such that the essential
spectrum is neither negative nor positive. This means that the
spectral subspaces both on the negative and positive side of the real
axis are infinite-dimensional. The topology in $\Cal F_*$ is defined
by the operator norm. The fact that Dirac operators are unbounded need
not bother us since we shall be really interested only on the spectral
resolutions defined by the sign operators $(D-\lambda)/|D-\lambda|,$
which are bounded. 

The space $\Cal F_*$ retracts onto the subspace $\hat\Cal F_*$
consisting of operators with essential spectrum at $\pm 1,$ [AS].
Thus it is sufficient to study the $U_{res}$ bundle $\Cal P'$ over 
$\hat \Cal F_*.$ The fiber $\Cal P'_A$ at $A\in\hat\Cal F_*$ is
defined as the set of unitary operators $g$ in $H_+\oplus H_-$ such
that $g\cdot H_+$ belongs to the Grassmannian $Gr(W_A)$ where $W_A$ is the 
positive spectral subspace of the operator $A.$ 

The proof of the local triviality of the bundle $\Cal P'$ is a slight
extension of the argument which was used in the proof of Proposition
1. First, we can choose an operator norm continuous (smooth) section 
$F:U(H)/(U(H_+)\times U(H_-))\to U(H)$ (again
by  Kuiper's theorem). Defining $U_{\lambda},$ for  $1> \lambda>-1,$ as
the open set in $\hat\Cal F_*$ consisting of operators $A$ such that 
$\lambda\neq Spec(A)$ then $g_{\lambda}(A)=
F((A-\lambda)/|A-\lambda|)$ is a local section of $\Cal P'$ with local
norm continuous transition functions $g_{\lambda\mu} =
g_{\lambda}^{-1} g_{\mu}.$ The Hilbert-Schmidt norm continuity (and
smoothness)  of the off-diagonal blocks 
follows as in Proposition 1 since, by our assumption about the
essential spectrum of $A$, the spectral subspaces corresponding to the
open intervals $]\mu,\lambda[$ are finite-dimensional for
$-1<\mu<\lambda<1.$  



\vskip 0.3in

3. A UNIVERSAL CONSTRUCTION

\vskip 0.3in

In this section $G$ denotes the group of unitary operators $g$
in $H$ such that $g-1$ is trace-class.

Let $\Cal P$ be a locally trivial principal $U_{res}$ bundle over $G.$
The Lie algebra of $\Ut$ is
defined by the standard 2-cocycle
$$c(X,Y) =\frac14 \tr\, \epsilon[\epsilon,X][\epsilon, Y].\tag1$$
Let $(\Cal U_{\alpha})$ be a family of open contractible sets covering
the base $G.$  Let $\Cal L\to U_{res}$ be the complex line bundle associated
to the circle bundle $1\to S^1\to \Ut \to U_{res} \to 1.$ Let $\phi_{\alpha}:
 \Cal U_{\alpha} \to \Cal P$ be a family of local trivializations of the
bundle $\Cal P.$ Let $g_{\alpha\beta}$ be the corresponding family of $U_{res}$
valued transition functions. We define a family of local line bundles
over the open sets $\Cal U_{\alpha\beta} =\Cal U_{\alpha}\cap \Cal
U_{\beta}$  by pull-back,
$\Cal L_{\alpha\beta}= {g_{\alpha\beta}}^* \Cal L.$

Since $\Ut$ is a group we have a natural isomorphism
$$\Cal L_g \otimes \Cal L_f \equiv \Cal L_{gf} \tag2$$
for all $g,f\in U_{res}.$  This gives a family of isomorphisms
$$\Cal L_{\alpha\beta} \otimes \Cal L_{\beta\gamma} = \Cal L_{\alpha\gamma}\tag3$$
over the common intersections $\Cal U_{\alpha\beta\gamma}=\Cal U_{\alpha}\cap
\Cal U_{\beta} \cap \Cal U_{\gamma}.$ Thus we have a
bundle gerbe $\Cal Q$ over the base $G,$ \cite{CMM1,CMM2}. 
The product structure gives also a natural isomorphism $\Cal L_g
\equiv {{\Cal L}_{g^{-1}}}^{-1}$ and therefore an isomorphism $\Cal
L_{\alpha\beta} \equiv {\Cal L_{\beta\alpha}}^{-1}.$ Combining this
with (3) we obtain a natural trivialization $f_{\alpha\beta\gamma}$ 
of the product bundle $\Cal L_{\alpha\beta}\otimes \Cal
L_{\beta\gamma} \otimes \Cal L_{\gamma\alpha}$ over
$\Cal U_{\alpha\beta\gamma}.$  

The family $\{f_{\alpha\beta\gamma}\}$ of trivializations (local $S^1$
valued functions) satisfies the
cocycle condition 
$$f_{\b\g\delta} f_{\a\g\delta}^{-1} f_{\a\b\delta}f_{\a\b\g}^{-1}=1$$ 
on the intersections of four open sets.

Because of the relations (3) the local curvature forms $\omega_{\alpha\beta}=
g_{\alpha\beta}^* c $ satisfy the relations
$$[\omega_{\alpha\beta}]
+[\omega_{\beta\gamma}]+[\omega_{\gamma\alpha}]=0\tag4a $$
in de Rham cohomology $H^2(\Cal U_{\a\b\g})$ on the base. Note that these equations do
not hold on the level of differential forms. However, this
can be corrected by adding an exact 2-form $d\theta_{\alpha\beta}$ to
each of the closed forms $\omega_{\alpha\beta};$ the modified forms
$\omega_{\a\b}$ satisfy then
$$\omega_{\a\b}+\omega_{\b\g} +\omega_{\g\a} =0. \tag4b$$ 
Actually, because of
the given local trivializations $f_{\a\b\g}$ on triple intersections
we have the consistency condition  
$$ \nabla_{\a\b\g} f_{\a\b\g}=0,\tag5$$ 
where $\nabla_{\a\b\g}$
 is the connection on the trivial bundle $\Cal L_{\alpha\beta}\otimes {\Cal
L}_{\beta\gamma} \otimes {\Cal L}_{\gamma\alpha}$  composed from the
connections on the factors ${\Cal L}_{\a\b}$ with curvature forms
$\omega_{\a\b}.$ In fact, (5) implies (4b): If $A_{\a\b}$ is a local
potential, $dA_{\a\b} =\omega_{\a\b},$  then (5) can be written as 
$$df_{\a\b\g} + (A_{\a\b} +A_{\b\g} +A_{\g\a})f_{\a\b\g} =0,$$   
and multiplying by $f^{-1}_{\a\b\g}$ and then taking the exterior
derivative gives the cocycle relations (4b) for the forms $\omega_{\a\b}.$

If $(\rho_{\alpha})$ is a partition
of unity on $G$ subordinate to the covering $(\Cal U_{\alpha})$ then
we can produce a closed 3-form on $G$ in the usual way.  First, we
have a closed 3-form $\omega_{\alpha}$ on $\Cal U_{\alpha},$
$$\omega_{\alpha} = d\sum_{\b}
\omega_{\a\b} \rho_{\b}
=\sum_{\b} \omega_{\a\b} d\rho_{\b} .
$$
 Since $\omega_{\a} -\omega_{\b}=0$ on $\Cal U_{\a\b}$ 
they can be
pasted together to give the closed 3-form $\omega_3$ on $G.$ 
This is the de Rham form of the 
Dixmier-Douady (DD) class of the bundle gerbe. Of course, this
description disregards all potential torsion information. However, for
our purposes the differential form picture is quite sufficient since
we going to show that there is already on this level an obstruction to quantization.

Unfortunately the existence of a
partition of unity on the infinite-dimensional manifold $G$ does not
appear to be known. However, in order to define the DD class as
an element of the dual of the
3-homology classes it is sufficient to use the partition of unity on
the (singular) homology 3-cycles and to pull-back the forms
$\omega_{\a\b}$ 
down to the embedded 3-cycles and then proceed
as above. An alternative solution would
be  to replace $G$ by the group of unitaries
differing from the identity by a Hilbert-Schmidt operator
(which is a Hilbert manifold) where we would
have a partition of unity. 

Note however that whichever method we use to define the DD class
we can normalize it
so that its integral
around closed 3-cycles is $2\pi$ times an integer.

In the case when the gerbe is coming from  a principal $U_{res}$
bundle $\Cal P$ over $G$ there is another method to construct the
Dixmier-Douady class which gives directly the integrals of $\omega_3$
over 3-cycles in $G.$ The homology cycles can be generated by mappings
from $S^3$ to $G,$ so we shall restrict to this case. 

Map the 3-disk $D^3$ onto the sphere $S^3$ such that the boundary is
mapped to one point $g\in S^3.$ Pulling back the bundle $\Cal P$ to 
$D^3$ leads to a trivial $U_{res}$ bundle over $D^3.$ In this
trivialization
the boundary $S^2\subset D^3$ is mapped to the fiber $\Cal P_g$ over
$g.$ Selecting a base point $x$ in the fiber we can identify $\Cal P_g
\simeq U_{res}.$ The integral of the curvature form $c$ over $S^2$
gives then the integral of $\omega_3$ over $S^3\subset G.$ Note that
the result does not depend on the choice of the base point $x$ since a
different choice $x'$ is related by a right translation $x'=x\cdot q$
by an element $q\in U_{res}.$ The cohomology class $[c]$ is invariant
under right (and left) translations on the group manifold.   One can
check that this construction gives the same result as the one starting
from the cocycle of 2-forms $\omega_{\a\b}.$ Namely, selecting a local
trivialization of the bundle $\Cal P$ at $g\in G$ such that $g$ is
mapped to the point $x$ in the fiber, we observe that the map $S^2 \to
U_{res}$ above is just the transition function from the local
trivialization over $D^3$ to the local trivialization around $g.$  The
rest follows from Stokes' theorem, using the fact that $\omega_{\a\b}
=d^{-1}\omega_{\a} -d^{-1}\omega_{\b}$ on the overlap and that
$\omega_{\a}=\omega_{\b}= \omega_3$ on $\Cal U_{\alpha\beta}.$

\noindent\bf Universal $U_{res}$ bundle over $G.$ \rm Let $\Cal P$ be the space
of smooth paths (parametrized by $0\leq t\leq 2\pi$) in $G$ with
initial point $1\in G.$ We also require that the derivatives of
$g(t)$ vanish at the end points.  To each $g\in \Cal P$ there corresponds a
vector potential $A$ on the circle $S^1,$ with values in the Lie
algebra $\bold{g}$ of $G,$  $A(t) = g(t)^{-1} g'(t).$ The group
$\Omega G$ of based loops at $1\in G$ acts freely from the right on
$\Cal P$ through gauge transformations $A^g= g^{-1} A g +g^{-1}g'$ and the
set of orbits is ${\Cal P}/\Omega G = G.$ Since clearly $\Cal P$
is contractible,
$\Cal P$ is the universal $\Omega G$ bundle over $G.$
The local triviality of the path fibration is obtained by the
exponential mapping; locally, near the unit element, the
trivialization is $(g,h)\mapsto \tilde h,$ where $g\in G,$ $h(t)$ is a
based loop at $1,$ and $\tilde h(t) = \exp(tZ) h(t)$ with $\exp(2\pi Z)=g.$

By Bott periodicity, $\Omega G$ is homotopy equivalent with $U_{res}.$ 
Actually, we can define a group homomorphism $j: \Omega G\to U_{res}$
which is a homotopy equivalence, see appendix 2 for proofs.
The group $G$ acts, by definition,
as the unitary group of $1$ + trace-class operators in a complex
Hilbert space $H.$   Let $\Cal H =L^2(S^1, H)$ and let $\Cal H=\Cal
H_+\oplus \Cal H_-$  be the polarization to nonnegative and negative
Fourier components. Then each $g\in\Omega G$ acts naturally in
$\Cal H$, by pointwise multiplication, and this action gives the
promised homomorphism to $U_{res}(\Cal H_+\oplus \Cal H_-).$  The homomorphism
can be used to define an associated bundle $\tilde{\Cal P}={\Cal P}
\times_{\Omega G} U_{res}$
with fiber $U_{res}.$ This latter bundle is then a universal $U_{res}$ bundle
over $G,$ see Proposition 2.

\noindent\bf Remark. \rm There is an alternative construction of the universal
$U_{res}$ bundle over $G$ which makes its appearance in
quantum field theory (QFT) more
transparent. For each gauge potential $A(t)=g(t)^{-1}dg(t), g\in \Cal P,$
 let $W_A$ be the set of unitary
operators $T$ in $\Cal H$ such that the projections onto the subspaces
$T\cdot \Cal H_+$ and $\Cal H_+(D_A)$ differ by a Hilbert-Schmidt
operator; here $\Cal H_+(D_A)$ is the positive frequency subspace for
the Dirac operator $D_A.$ It is easy to see that if both $T,T'\in W_A$
then $T'=Th$ for some $h\in U_{res}.$ Consequently, one can view $W_A$
as a fiber in a principal $U_{res}$ bundle over ${\Cal P}$ [CMM2]. The base
is contractible, thus there exists a global section $A\mapsto T_A.$
Actually,
we can construct explicitly a global section by setting
$$T_A= T_{f^{-1}df}= f,$$
that is, $T_A$ is the multiplication operator by the function $f(t)$
on the interval $[0,2\pi].$

We can define a $U_{res}$ valued 1-cocycle for the natural $\Omega G$
action on $\Cal P$ by
$$S(A;g) = T_A^{-1} T_{A^g} .$$
In the case of the choice $T_A=f$ above, we have simply $S(A;g)= g.$
This cocycle defines the same  associated  $U_{res}$ bundle $\tilde{\Cal P}$
over  $G$ as above, by the
equivalence relation $(A,T) \simeq (A^g,T S(A;g))$ for $g\in\Omega G$
and $T\in W_A.$

\proclaim{Proposition 2} The total space of the bundle $\tilde P$ is
contractible and thus $\tilde P \to G$ is the universal $U_{res}$
bundle over $G.$ \endproclaim
 \demo{Proof} By a well known theorem in homotopy theory \cite{G} a
space with the homotopy type of a CW complex
is contractible if it is weakly homotopy equivalent to a point.
All of the spaces we consider are Banach manifolds
and have the homotopy type that of a CW complex. We want to compare
the homotopy groups of $\Cal P$, which are all trivial, with the homotopy
groups $\pi_i(\tilde {\Cal P}).$ From
 the appendix 2 we know that the embedding $j$
of $\Omega G$ to $U_{res}$ is a homotopy equivalence.
This embedding extends to an embedding of
principal bundles $\tilde j: {\Cal P}\rightarrow \tilde {\Cal P}$
by mapping $p\in \Cal P$ to the equivalence class
of the pair $(p,e)\in {\Cal P} \times U_{res}$ in $\tilde {\Cal P}$; $e$
is the identity element in $U_{res}.$ 
On each fibre this map reduces to $j$.
This means we have
a commutative diagram
$${\matrix \pi_i({\Omega G})&{\buildrel {j^*} \over \longrightarrow}
&\pi_i(U_{res})\cr
\ \ \ \nwarrow&  &\nearrow\ \ \ \cr
   &\pi_{i+1}(G)&  \cr \endmatrix}$$
where the up arrows
represent the connecting maps for the homotopy long exact
sequences for the locally trivial fibrations $\Cal P$
and $\tilde {\Cal P}$ respectively.

Since $\pi_i(P)=0$ for all $i$, by the homotopy exact sequence

$$...\pi_{i+1}(G)\to \pi_i(\Omega G) \to \pi_i(P)\to \pi_i(G) \to...$$
the connecting map $k: \pi_{i+1}(G) \to \pi_i(\Omega G)$ is an
isomorphism (both groups are known to be $0$ when $i$ is odd and equal
to $\Bbb Z$ when $i$ is even (these facts rely on results
of Palais, see \cite{Q} for a discussion). From the commutative
diagram above it follows that the corresponding map for the fibering
$U_{res} \to \tilde P \to G$ is also an isomorphism and from the 
homotopy long exact sequence for this fibering we conclude that
$\pi_i(\tilde P)=0.$ Moreover this shows
that $\tilde j$ is a weak homotopy equivalence
from which we deduce that $\tilde P$ is a contractible.  \enddemo

\vskip 0.3in
4. THE DIXMIER DOUADY CLASS

\vskip 0.3in

Next we want to relate the Dixmier-Douady class $\omega_3\in H^3(G,
\Bbb Z)$ of the various gerbes over $G$ to the
natural curvature 2-form on the group
$\Omega G.$ Recall that this curvature form on $\Omega G$ is the homogeneous
2-form
$$\omega_2= \frac{1}{2\pi}\int_{S^1} \tr\, (g^{-1}dg)\frac{d}{dt} (g^{-1}dg).\tag6$$

We start from the universal $\Omega G$ bundle $\Cal P$ over $G$ and
its bundle gerbe as described in Section 2.

\proclaim{Proposition  3} The cohomology class $[\omega_3]$ is represented
by the closed 3-form

$$\theta_3= \frac{1}{12\pi} \tr\, (g^{-1}dg)^3.\tag7$$
\endproclaim

\demo{Proof} We have to show that the integral of $\theta_3$ over any closed 3-cycle
on $G$ agrees with the integral of $\omega_3.$ We can generate
3-cycles by 3-spheres, so we take as the 3-cycle a (differentiable)
mapping $\chi$ of $S^3$ into $G.$ Since $G$ is connected, we can
assume
the image under $\chi$ of $S^3$ is
such that the poles of $S^3$ are mapped to the unit element in $G.$

We cut the 3-cycle $S^3$ along the equator to two disks $B_+$ and $B_-.$
Over these disks the pullback under $\chi$ of the
principal bundle ${\Cal P} \to G$ is trivial; choose a pair
of local trivializations $\psi_{\pm}$ over $B_{\pm}.$
Concretely, the local trivializations can be defined as follows. For
each $x\in S^2$ on the equator we have path $t\mapsto \phi_+(x)(t)\in S^3$
by connecting the point $x$ by a segment of a great circle to the to
the 'north pole'. Similarly, we obtain a path $\phi_-(x)(t)$ by
connecting $x$ by a great circle to the 'south pole'. We choose the
parametrization such that $t=2\pi$ corresponds to the point $x$ and $t=0$
corresponds to either of the poles (which is mapped to the unit
element in $G$). Setting $\psi_{\pm}=\chi\circ \phi_{\pm}$ the
transition  function on the equator (with values in
$\Omega G$) is then given by $\psi_+(x,t) =\psi_-(x)(t) g(x,t)$
for some $g(x,.)\in \Omega G$.  On the other hand,
$(x,t)\mapsto\psi_{\pm}(x,t)$ is a parametrization of points on
$\chi(B_{\pm}) \subset G$ and therefore, by a simple
calculation,

$$\align 12\pi \int_{S^3} \theta_3 &= \sum_{\a=\pm}\int_{B_{\a}} \tr\,({\psi_{\a}}^{-1}
d\psi_{\a})^3\\
&=\int_{S^2\times S^1} \tr\, (g^{-1}dg)^3 +
3 \int_{S^2} \tr\, {\psi_-}^{-1} d\psi_-\wedge dg g^{-1}.\endalign$$

The second term on the right vanishes since $g(x,2\pi)=1$ on $S^2.$   The first
term on the right is

$$\int_{S^2\times S^1} \tr\,(g^{-1}dg)^3 = 12\pi \int_{S^2} \omega_2. \tag8$$

On the other hand, the integral on the right is equal to the integral of
$\omega_3$ over $S^3;$ this follows, by Stokes' theorem,  directly from the construction of the
DD class $\omega_3$ from the family of local 2-forms $\eta_{\a}$
such that $\eta_{\a} -\eta_{\b} = \omega_{\a\b}$ and $d\eta_{\a}
= \omega_{\a}=\omega_3|_{U_{\a}};$ recall that the class $[\omega_{\a\b} ]$ is
given by the pull-back of $\omega_2$ with respect to the transition
function $g: \Cal U_{\a\b} \to \Omega G.$ 

Thus indeed
$$\int_{S^3} \theta_3 = \int_{S^3} \omega_3\tag9$$
and so $\omega_3$ and $\theta_3$ represent the same cohomology class.
\enddemo

The DD class is an obstruction to writing
the line bundles $\Cal L_{\a\b}$
as tensor products $\Cal L_{\a}\otimes \Cal L_{\b}^*$ of local line bundles over
the open sets $\Cal U_{\a}$ on the base, that is, an obstruction to a
trivialization of the gerbe $\Cal Q.$

We use the following theorem, taken from [\cite{K}, 
Theorem 3.17] in a slightly reformulated form:

\proclaim{Theorem 1} [K]. Let $M$ be a compact space  and $[M, G]$
the set of homotopy classes of maps from $M$ to $G.$ Then
$K^1(M)$ is isomorphic with $[M,G].$ The group structure in the
latter group is given by pointwise multiplication of maps. \endproclaim


On the other hand, we have seen that $\Cal P$ can be viewed as an universal
$U_{res}$ bundle (or, what is essentially the same, as an $\Omega  G$ bundle)
over $G.$ Thus we can construct $U_{res}$ bundles
over any compact space $M$ by pulling back this universal bundle
via a map   from $M$ to $G$.

The index theorem in [MP], Proposition 12,  tells us  that
for a compact subset $M\subset G$  $K^1(M)$ is
realised by homotopy classes
of maps into the family of odd-dimensional Dirac operators
parametrized by the boundary conditions $M.$
By  the above theorem
we can conclude that the universal $U_{res}$ bundle over $G$ can
be identified as a (universal) $U_{res}$ bundle over the family of
Dirac operators (identified topologically as the family $G$ of boundary
conditions).

In the case of the universal $U_{res}$ bundle  over $G$ we already know the DD class
of the bundle gerbe $\Cal Q$ is given by the generator (7) (devided by
$2\pi$)  of $H^3(G, \Bbb Z).$
Thus this is also the (nontrivial) obstruction to a trivialization of
the gerbe over the space of Dirac operators parametrized by the
boundary conditions $G.$

As a conclusion we obtain our main result:

\proclaim{Theorem 2} There is an obstruction to a prolongation of the
$U_{res}$ bundle $\Cal P$ (as defined in  Section 2)
over a compact submanifold $M\subset G$ of hermitean elliptic
boundary conditions, that is, an obstruction to the construction of the
bundle of fermionic Fock spaces for the Dirac operators parametrized
by $M.$ The Dixmier-Douady class of the obstruction is given by the
restriction of the de Rham class $\theta_3$ to $M.$ \endproclaim

In particular, the obstruction is nontrivial when $M=U(N)$ is any
finite-dimensional subgroup of $G$ with $N\geq 2.$

\noindent{\bf Remark 1}. We could
interpret our family of
Dirac operators parametrised by $G$ as  an element of
$K^1(G)$ if the latter were defined as
homotopy classes of  maps into the self adjoint Fredholm operators.
As $G$ is not compact this is problematic.
For our purposes it is enough to work with compact subsets.

\noindent {\bf Remark 2}. In the case of an odd-dimensional manifold
without boundary there is a similar obstruction to \it gauge invariant
\rm quantization, related to Schwinger terms in current algebra,
[CMM1, CMM2]. Recently the case of gravitational Schwinger terms was
discussed in the same formalism, [EM].

\vskip 0.3in

APPENDIX 1: PROOF OF THE PROPOSITION 1

\vskip 0.3in 

Let
$T_0$ be any fixed boundary condition, with $T_0:S^+\to S^-$ a unitary
map. If $T$ is another boundary condition then the graph of $T$ is
obtained from the graph of $T_0$ by the unitary transformation
$(u_+,u_-)\mapsto (u_+,g\cdot u_-)$ with $g=TT_0^{-1} \in G$  and
$u_{\pm} \in S^{\pm}.$
In a small open
neighborhood $U$ of $1\in G$ we can choose in a smooth way, for any $g\in U,$  a smooth
path $g(t)$ such that $g(0)=1$ and $g(1)=g;$ this is achieved for
example by writing $g=\exp(Z)$ and
setting $g(t)=\exp(tZ).$

Near the boundary $Y$
we may think of the $L^2$ functions on $X$ as functions $f(t,y)$ on $[0,1]
\times Y$ (where $t$ is a parameter in the normal direction at the
boundary) and we can extend the action of $g$ in the boundary Hilbert
space $L^2(Y,S)$ to an action on $H=L^2(X,S)$ by setting
$$(R(g) f)(t,x)= (f_+(t,x) ,g(t)\cdot f_-(t,x))$$
in the tubular neighborhood $[0,1]\times Y$ and $R(g)$ acts as an identity
on $f$ outside of this neighborhood. The map $g\mapsto R(g)$ is
smooth  with respect to the $L^1$ norm in $G$ and the operator norm
in the algebra of bounded operators in $L^2(X,S).$ This follows from
the smoothness  of the embedding of trace-class operators to the
algebra of bounded operators and from the smoothness of the
exponential mapping.

Clearly $R(g)$ is an unitary operator in $L^2(X,S)$ and it maps the
domain $dom(D_g)$ onto the domain $dom(D_1)$ of the reference Dirac
operator. Thus $D_g$ is unitarily equivalent to the Dirac operator
$R(g)D_g R(g)^{-1}$ in the fixed reference domain $dom(D_1).$

Next we choose a smooth  mapping from the space of polarizations
$\epsilon$ in
$H$ to the unitary group $U(H)$ such that $\epsilon= F(\epsilon)
\epsilon_0 F(\epsilon)^{-1}$ where $\epsilon_0$ is the fixed reference
polarization given by the sign of the Dirac operator $D_1.$ This
mapping exists because the space of polarizations $U(H)/(U(H_+)\times
U(H_-))$ is a contractible Banach manifold (with respect to the
operator norm) by Kuiper's theorem.

With
these tools we can write an explicit local trivialization of the
Grassmannian bundle $Gr$ over $G.$ Near the unit element in $G$  the
Hilbert-Schmidt Grassmannians are parametrized as
$$ (g,\epsilon)\mapsto F(\epsilon_g) \epsilon F(\epsilon_g)^{-1},$$
where $\epsilon\in Gr_1$ and $\epsilon_g= R(g) \frac{D_g}{|D_g|}
R(g)^{-1}.$ We have assumed that zero does not belong to the spectrum
of $D_g;$ otherwise, we replace $D_g$ by $D_g -\lambda$ for some real
number $\lambda$ in the neighborhood of $g=1.$

Because of the potential non-triviality of the
kernel of the operator $D_g$ we cannot have a global
trivialization of the bundle. However, for each real number $\lambda$ the
trivialization described above is well-defined in the open set $G_{\lambda}$
consisting of those elements $g$ for which $\lambda\notin Spec(D_g).$
The transition function on the overlap $G_{\lambda}\cap G_{\mu}$ is then
given by
$$\epsilon \mapsto F(\epsilon_g(\lambda))^{-1} F(\epsilon_g(\mu))\epsilon
F(\epsilon_g(\mu))^{-1} F(\epsilon_g(\lambda)),$$
where $\epsilon_g(\lambda)$ is defined as $\epsilon_g$ above but with the
shifted operator $D_g - \lambda.$ By the construction, the transition
function
$h_{\mu\lambda}(g)= F(\epsilon_g(\mu))^{-1} F(\epsilon_g(\lambda))$
satisfies
$$[\epsilon_0, h_{\mu\lambda}] =
F(\epsilon_g(\mu))^{-1}\Delta_{\mu\lambda} F(\epsilon_g(\mu))$$
with $\Delta_{\mu\lambda}=\epsilon_g(\mu) -\epsilon_g(\lambda).$ Now
on the overlap $G_{\lambda}\cap G_{\mu}$ the difference
$\Delta_{\mu\lambda}$ has constant finite rank and therefore also
$[\epsilon_0, h_{\mu\lambda}]$  has constant finite rank. A norm
continuous mapping to operators of constant finite rank is continuous
also with respect to the Hilbert-Schmidt norm (or with respect to any
$L_p$ norm) which proves the continuity of the $U_{res}$ valued transition
functions. The same argument can then be used for the derivatives of
the transition function.

\vskip 0.3in

APPENDIX 2: PROOF OF THE HOMOTOPY EQUIVALENCE $j$

\vskip 0.3in

We define a system of closed $n$ forms $(n=2,4,6,\dots)$ on the
Hilbert-Schmidt Grassmannian $Gr(H_+)$ for the polarization
$H=H_+\oplus H_-,$  by
$$\omega_{n} = a_n \tr\, F (dF)^n,\tag A-1$$
where $F$ is the grading operator associated to $W\in Gr(H_+),$ that
is, $F$ restricted to $W$ is $+1$ and the restriction to $W^{\perp}$ is
$-1.$ Note that since $F^2=1,$ the differentials $dF$ anticommute with
$F.$ Thus for odd $n$ the form $\omega_n$ vanishes identically; $a_n$
is a normalization coefficient given by
$$a_n = - (\frac{1}{2\pi i})^j \frac{(j-1)!}{(2j-1)!}, \text{ with }
n=2j.$$
With this normalization the form $\omega_n$ is the generator in
$H^n(Gr(H_+), \Bbb Z).$


Since $U_{res}/(U(H_+)\times U(H_-)) = Gr(H_+)$ and the diagonal
subgroup is contractible, the natural projection $p: U_{res}\to
Gr(H_+)$ can be used to pull back the generator $\omega_n$ to the
generator $\phi_n$ in $H^n(U_{res}, \Bbb Z).$ The projection can be
written as $p(g) = g\epsilon g^{-1} =F_g,$ where $\epsilon$ is the
grading associated to $W=H_+.$  It follows that the pull-back of $dF$
is $[\theta, F_g],$ where $\theta= dg g^{-1}$ is the Maurer-Cartan
1-form on $U_{res},$ and so
$$\phi_n = a_n \tr\, F_g [\theta, F_g]^n.\tag A-2$$

The homotopy type of both $\Omega G$ and $U_{res}$ is known. The
homotopy groups vanish in odd dimensions whereas in even dimensions
the homotopy groups are all isomorphic with $\Bbb Z,$ \cite{C},\cite{PrSe},
\cite{Q}.
A generator $x_n$
of the homotopy group $\pi_n(U_{res})$ when paired with the generator
$\phi_n$ of $H^n(U_{res},\Bbb Z)$ gives $<x_n,\phi_n>=1.$ Thus the only
thing we need to check to prove that the embedding $j:\Omega G \to
U_{res}$ is a homotopy equivalence is to show that $<j(y_n), \phi_n>
=1$ for all even $n,$ where $y_n$ is the generator of $\pi_n(\Omega G).$

The odd generators in $H^*(G,\Bbb Z)$ are given by the differential forms
$$\psi_{2j-1} =  a_{2j} \tr\,
\theta^{2j-1}.\tag A-3$$

{}From this one obtains by transgression the even generators in
$H^*(\Omega G, \Bbb Z),$
$$\psi'_{2j} = (2j+1) a_{2j+2}   \int_{S^1} \tr\,
(g'(t) g(t)^{-1}) \theta^{2j}.\tag A-4$$

We shall show that $\psi'_n$ is in the same cohomology class as the
restriction of $\phi_n$ to the subgroup $\Omega G \subset U_{res}.$
Provided that this is the case, we have $1=<y_n, \psi'_n> =
<j(y_n),\phi_n>$ and thus indeed $j(y_n)$ is the generator of
$\pi_n(U_{res}),$ and consequently $j$ is a homotopy equivalence.

Note that $[\theta, F_g]= g[g^{-1}dg, \epsilon] g^{-1}$ and therefore
$$\phi_n = a_n \tr\, \epsilon [\theta_L, \epsilon]^n,\tag A-5$$
where $\theta_L = g^{-1} dg.$ The proof of the equivalence of $\phi_n$
and $\psi'_n$ is through a standard trick using the Cartan homotopy
method. We set $\theta_L = \theta_0 + \theta_1$ where $\theta_0$
commutes with $\epsilon$ and $\theta_1$ anticommutes with $\epsilon.$
Define $G_t = -t \theta_1^2 + (1-t) \theta_0^2$ for $0\leq t\leq 1.$
Then by $d\theta_L = -\theta_L^2$ we get

$$\align d G_t &= -\frac{1}{t} [\theta_0, G_t] \\
         \frac{d}{dt} G_t &= -(\theta_0^2 + \theta_1^2) =
d\theta_0.\tag A-6\endalign$$
Thus we obtain

$$ \tr\, \epsilon G_1^n  -\tr\,\epsilon G_0^n = \int_0^1 \tr\,\epsilon
\frac{d}{dt}
G_t^n dt = -n \int_0^1 \tr\,\epsilon (\theta_0^2 +\theta_1^2) G_t^{n-1}dt =
n d \int_0^1 \tr\, \epsilon \theta_0 G_t^{n-1} dt.\tag A-7$$

Actually, $\epsilon G_0^n$ is not trace-class. However, this operator
is an even wedge power of  $\theta_0$ and by the permutation
properties of the wedge product it is a commutator of some zero order
pseudodifferential operators on the circle. Such an operator is always
conditionally trace-class: this means that the trace is evaluated by
first taking the trace over matrix indices, then integrating the
product symbol over the circle and finally integrating over the
momentum variable from $-\Lambda$ to $\Lambda$ and taking the limit
$\Lambda\to \infty.$ But for conditionally trace-class operators
$\tr [A,B] \neq 0,$ in general.

In the present case we can write
$$\tr\, \epsilon G_0^n = \frac12\tr\,
[\theta_0,\epsilon\theta_0^{2n-1}]\tag A-8$$
where we have used the fact that $\theta_0$ commutes with $\epsilon.$
If $a(x,p)$ and $b(x,p)$ are the symbol functions of a pair of zero order
PSDO's $A,B$
on the circle, then the conditional trace of the commutator $[A,B]$ is
given by the expression
$$\tr [A,B]= \frac{-i}{2\pi} \lim_{\Lambda\to\infty}
\int_{p=-\Lambda}^{\Lambda}dp
\frac{\partial}{\partial p}\int_x dx\,\tr\, \left(\frac{\partial a}
{\partial x}(x,p) b(x,p) - a(x,p) \frac{\partial b}{\partial x}
(x,p) \right).\tag A-9$$
Since in the expression (A-8) above the momentum variable is contained only
in $\epsilon = p/|p|$ and the derivative of this symbol is twice the delta
function in momentum  space, we obtain

$$\tr\, \epsilon G_0^n = \frac12 \frac{-i}{2\pi}\int_x \tr\, \theta^{2n-1}
\frac{d}{dx} \theta dx.\tag A-10$$

Note that in the case $n=1$ this is just the curvature on a loop group
arising from the canonical central extension of the loop algebra.

By integration in parts, we can conclude that the integral of (A-10) over
any compact $2n$ manifold $M_{2n}$ in $G$ is equal to the integral
$$\frac{-i}{4\pi}\int_{M_{2n}}\int_{S^1}dx\, \tr\, \theta^{2n} (g^{-1} \frac{d}{dx} g)=
\frac{-i}{4(2n+1)\pi}\int_{S^1 \times M_{2n}} \tr\,\theta^{2n+1} \tag A-11$$
and therefore this last expression under the integral sign represents the
same cohomology
class as (A-10), and therefore after a multiplication by the normalization
coefficient $a_{2n}$, through (A-7), the same class as (A-4). Note that the
coefficient in front of the integral on the right-hand side of the equation
is equal to the ratio $a_{2n+2}/a_{2n}$ which is necessary to recover the
correct normalization for $\theta^{2n+1}.$
This proves that $<j(y_n),\phi_n>=<j(y_n),\psi'_n>=1.$

\vskip 0.3in
\bf References \rm

[AS] M.F. Atiyah and I.M. Singer: Index theory for skew-adjoint Fredholm
operators. I.H.E.S. Publ. Math. \bf 37, \rm p. 305-326 (1969)

[C] A.L. Carey:
Infinite dimensional groups and quantum field theory  Acta
Applicandae Mathematicae, {\bf 1} (1983), 321-331.

[CMM1] A. L. Carey, J. Mickelsson, and M. Murray: Index theory, gerbes,
and hamiltonian quantization. Commun. Math. Phys. \bf 183, \rm 707
(1997). hep-th/9511151

[CMM2] ------:  Bundle gerbes applied to field
theory. hep-th/9711133. Rev. Math. Phys. \bf 12, \rm (2000), 65-90.

[EM] C. Ekstrand and J. Mickelsson: Gravitational anomalies, gerbes,
and hamiltonian quantization. hep-th/9904189. To be publ. in Commun. 
Math. Phys.

[G] B. Gray: Homotopy Theory: an introduction to algebraic topology,
Academic Press, New York, 1975.

[K] M. Karoubi: \it K-Theory. An Introduction. \rm Grundlehren der
Mathematischen Wissenschaften 226. Springer-Verlag, Berlin (1978)

[MP]  R.B. Melrose and P. Piazza, An index theorem for families of
Dirac operators on odd-dimensional manifolds with
boundary. J. Diff. Geometry \bf 46, \rm p. 287-334 (1997)

[PrSe] A. Pressley and G. Segal: \it Loop Groups. \rm Clarendon Press,
Oxford (1986)

[S] S. G. Scott:  Determinants of Dirac boundary value  problems over odd
dimensional manifolds. Commun. Math. Phys. \bf 173, \rm p. 43-76
(1995). Splitting the curvature of the determinant line
bundle. math.AP/9812124. 
To appear in Proc. Am. Math. Soc.

[W] E. Witten: Anti-de Sitter space and holography, Adv. Theor. Math. Phys.
{\bf 2} 253 (1998); hep-th/9802150.

[Q] D. Quillen:  Superconnection character forms
and the Cayley transform, Topology \rm {\bf 27} 1988, 211-238 and
P de la Harpe: \it Classical Banach-Lie algebras and Banach Lie groups
of operators in Hilbert Space, 
\rm Springer Lecture Notes in Mathematics {\bf 285} 1972

\enddocument